# Evolutionary conservation of motif constituents within the yeast protein interaction network

S. Wuchty[1], Z.N. Oltvai[2] and A.-L. Barabási[1]

[1]Department of Physics, University of Notre Dame, Notre Dame, IN 46556, USA

[2]Department of Pathology, Northwestern University, Chicago, IL 60611, USA

**Understanding why some cellular components are conserved across species, while others evolve rapidly is a key question of modern biology[1-3]. Here we demonstrate that in *S. cerevisiae* proteins organized in cohesive patterns of interactions are conserved to a significantly higher degree than those that do not participate in such motifs. We find that the conservation of proteins within distinct topological motifs correlates with the motif's inter-connectedness and function and also depends on the structure of the overall interactome topology. These findings indicate that motifs may represent evolutionary conserved topological units of cellular networks molded in accordance with the specific biological function in which they participate.**

Many biological functions are carried out by the integrated activity of highly interacting cellular components, referred to as functional modules[4,5]. Motifs, considered as topologically distinct interaction patterns within complex networks, may represent the simplest building blocks of such modules[6,7]. Owing to their small size, motifs can be explicitly identified and enumerated in various cellular networks[6-8], but their biological significance, if any, remains undetermined. A well-known signature of the conservation of specific cellular functions is the evolutionary retention of orthologous proteins that are responsible for selected functions. Therefore, the tendency to evolutionarily conserve the protein components of topologically distinct motifs could provide strong evidence for their importance and involvement in the definition of specific biological functions.

To test the correlation between a protein's evolutionary rate and the structure of the motif it is embedded in, we first identified all two to four-node, and some of the five-node motifs in the protein interaction network of *S. cerevisiae* utilizing the DIP protein interaction database[9]. Although the quality of two-hybrid results, supplying the core of the data, is a matter of debate[10], the manually curated DIP database represents our current best approximation for yeast protein interactions, providing sufficient data for their unambiguous statistical analyses (see the Supplementary Note for details). We find that the network of 3,183 interacting yeast proteins encodes between $10^3$ to $10^6$ copies of the specific motif types (Table 1).

If there is indeed an evolutionary pressure to maintain specific motifs, their components should be evolutionarily conserved, having an identifiable ortholog in other organisms. To test this hypothesis, we studied the conservation of the *S. cerevisiae* proteins by identifying a list of 678 proteins that have an ortholog in each of the five higher eukaryotes we use in our study, as deposited in the InParanoid database[11]. We find significant differences between the conservation rate of proteins within the different motifs: while less than 5% of the linear three node motifs (#2, Table 1) are completely maintained (i.e., all three component proteins have an ortholog), 47% of the fully connected pentagons are completely conserved across each of the other five eukaryotes (#11, Table 1).

These results indicate that the orthologs are not randomly distributed in the yeast protein interaction network, but are the building blocks of cohesive motifs, which tend to be evolutionary conserved. We need, however, to test the validity of this finding against a random set of orthologs. If the same number of orthologs were randomly placed on the yeast protein interaction network, mimicking the absence of correlations between the network topology and the ortholog position, the motif conservation observed above should disappear. Indeed, we find that motifs under such random ortholog distribution (see Methods) display a trend opposite to that observed for the original non-random system: the larger the motif, the smaller is the likelihood that each of its components is conserved (Table 1). For example, while 4.6% of the randomized two node motifs (#1) are retained with randomized orthologs, this fraction is only 1.01% for the triangle (#3) -, 0.08% for the fully connected square (#9) -, and 0.02% for the fully connected pentagon motifs (#11).

The influence of the global network topology on the retention rate of specific local motifs is best quantified by calculating the ratio between the real- and the random conservation rate. We find that for each motif this conservation ratio is bigger than one, and increases significantly for larger motifs. Indeed, while for the two node motifs (#2) the conservation ratio is 2.94, for the larger fully connected motifs, such as the triangle- (#3), and square motifs (#9) it increases to 20.28, and 422.78, respectively. Moreover, the conservation rate of proteins participating in fully connected pentagon motifs is 2,256 times higher than expected if the network topology does not influence the natural placement of orthologs (#11, Table 1).

We also observe that larger motifs have a tendency to be conserved as a whole, each of their components having an ortholog. For example, less than 1% of the fully-connected pentagon motifs disappear completely, such that none of their protein components are conserved in other eukaryotes, and less than 2% of such pentagons have only one conserved protein (Fig. 1b). In contrast, for 69% of the fully-connected pentagons each of the subunits has an ortholog in humans. A similar trend toward complete conservation of larger motifs is observed for each of the five higher eukaryotes (Fig. 1a,b). In general, with increasing number of nodes within a motif and number of links among its constituents, the evolutionary retention of the constituent proteins is increasingly complete. In particular, we observe a clear correlation between the conservation rate and the motif's degree of saturation. Considering the four-nodes motifs, the more intraconnected ones (# 8, 9, Table 1) have a much higher conservation rate than their less intraconnected counterparts (#4, 5, 6, 7, Table 1). Overall, these exceptionally high conservation rates strongly suggest that participation within motifs significantly influences the evolutionary conservation of their specific components.

To examine the relationship between the network's local interconnectedness and the protein components' retention rate, we also measured the correlation between the clustering coefficient and the conservation rate of the interacting proteins (Fig. 1c). The clustering coefficient is high ($C_i = 1$) in a highly cohesive region of the network if all neighbors of a protein $i$ have links to each other, and is small ($C_i = 0$) if the network is locally sparse[12,13]. As Fig. 1c indicates, 65% ($C = 0$) to 84% ($C = 1$) of neighbors of a human ortholog are also human orthologs, the conservation rate increasing with the neighborhood's cohesiveness. In contrast, the conserved fraction of the non-orthologous protein's neighborhood is significantly smaller, decreasing from 40% ($C = 0$) to 20% ($C = 1$) (Fig. 1c). Therefore, groups of proteins forming a highly interlinked cluster tend to be (non)conserved in a cohesive group if they represent an evolutionary (non)conserved functional module.

Motifs and the clustering coefficient probe the network's small-scale properties, addressing the influence of a protein's immediate neighbors on its conservation rate. Yet, the proposed hierarchical modularity of metabolic-[13] and protein interaction networks[14], suggests that highly interconnected motifs may combine into larger, less cohesive modules. To examine if the observed correlations between the conservation rate and the network's topology are relevant beyond the protein's immediate vicinity, starting from ortholog $i$ we identified all proteins that are $d = 1,2,3,...$ links from $i$, where $d$ represents the shortest distance between $i$ and a target protein measured along the network links. We separately determined the fraction of orthologous proteins at distance $d$ for both the natural and the random ortholog distribution. The ratio of the natural and the random fractions of orthologous proteins, shown in Fig. 1d, indicate a considerable enrichment for orthologs at distances $d = 1,2,3$, which, however, disappear for $d > 3$. Indeed, we find that proteins which interact directly with an ortholog at $d = 1$ have a 50% or higher chance of conservation than expected for a random ortholog distribution, while those at $d = 2$ links away have a 25-35% higher rate of conservation. Enrichment (20-25%) is also observed for $d = 3$, indicating that the extended vicinity of an orthologous protein is enriched with orthologs, thus supporting the extension of conservation to larger modules as well.

To examine if the specific function of the yeast proteins within motifs affects their rate of evolutionary conservation we assigned each motif to the functional class to which all of its protein components belong, utilizing the classification of the MIPS database[15]. The results indicate that larger motifs display a remarkable functional homogeneity. Indeed, for 95% of those fully connected yeast pentagon motifs (#11) whose proteins have an ortholog in each of the five higher eukaryotes all components share at least one common functional class. In contrast, only 10% of the two node motifs (#1) are functionally homogenous. We identified the type and number of evolutionary fully conserved motifs of each functional class in *S. cerevisiae*, limiting our study to those proteins that have an ortholog in humans. The ratio of the number of motifs identified for the natural- and random ortholog distribution indicates significant, functional class dependent differences between the evolutionary conservation of motifs (Table 2). For instance, we find that in three functional classes (subcellular localization, protein fate and transcription) each of the eleven studied motifs are significantly over-represented. In contrast, a few functional classes, such as transport facilitation, regulation and cellular transport, have only one or two characteristic motifs and others have none. These results indicate that the different functions are not only associated with characteristic topological motifs, but that they also conserve these motifs at different rates during evolution.

While motifs may represent various types of protein interactions, the fully connected motifs (#9, #11), as expected, have a tendency to identify protein complexes. Indeed, smaller complexes, in which each of the proteins interacts with all others should appear as fully connected *n*-node motifs in the protein interaction network. However, in larger protein complexes not all proteins have direct interactions with each other, thus motifs are expected to capture only some local, physically interacting components of the whole complex. For example, proteins found in the fully connected pentagons contain components of known yeast proteasome complexes RPN (rpn1,2,3,4,6,7,9,A,C), PSA (psa1,2,3,4,6,7), PSB (psb2,3,4,5,6) and PRS (psa4,6,7,8,A). These complexes are highly interacting with each other as well as with seven other proteins (sug2, mpr1, ra23, ubp6, and pyrg, p2a2, psda) that are not known to be part of the specific complexes. A separate cluster of proteins, in contrast, does not represent a protein complex, but consists of a highly interlinked collection of nucleolar (nop2,nop4, nog1), kinase (kc21) and RNA helicase (mak5, has1) proteins and four proteins with unknown functional role (ymt9, ytm1, yo26, yev6). The high number of interactions with these uncharacterized proteins may indicate functional relatedness, suggesting that a combination of evolutionary retention and dense interactions, as selected by the specific motifs, could be used to predict *in silico* the functional role of the unknown protein components. Note, however, that the mere existence of protein complexes cannot explain the observed trends towards increased conservation rate of the highly connected motifs. Indeed, we find that the basic conservation trends are not altered after the removal of the proteins that are part of known complexes, albeit the actual conservation ratios are changed. Similarly, although the protein interaction and orthologue databases are incomplete and contain numerous false positives, an error analysis confirms that our main findings and conclusions are not affected by such data inconsistencies, indicating the robustness of the observed evolutionary trends (see the Supplementary Note for details).

Further studies on the evolutionary conservation of topological modules and motifs would benefit from the simultaneous study of the retention rate of both nodes (e.g., proteins) and the links (interactions) among them. As among all eukaryotes protein interactions are available systematically only for *S. cerevisiae* our study is limited to the orthologous retention of the protein components of selected motifs. The high retention rate of many of the constituents of highly connected motifs (Table 1) strongly suggest that the interactions between the proteins of these motifs may be preserved in other organisms, a hypothesis that could be confirmed once protein interaction databases are established for other eukaryotic species.

Previous results suggest that the evolutionary rate of a protein correlates with the protein's essentiality and individual fitness[16-18] and its level of interactions with other proteins[19], but the quantitative correlations supporting some of these hypotheses were occasionally questioned[16,20,21]. As these hypotheses aim to relate the properties of cellular components to their evolutionary rate, the contradictory nature of some of these conclusions might have biological origins. Natural selection is expected to preserve components only to the degree they contribute to conserved cellular functions. Yet, a given biological function can be rarely assigned to a single protein, gene or metabolite, but emerges from the interaction of many separate components forming distinct functional modules[4,5,22]. Thus, the uncovered motif conservation may represent the network equivalent of domain and residue conservation in protein sequences. Our results indicate that understanding the evolutionary rate of single proteins must address the need to evolutionary preserve the specific functional modules, and the topologic features of the network their respective proteins are embedded in. In agreement with this hypothesis, we find that the conservation rate of motif constituents is increased by factors that range from tens to thousands, an enhancement that is clearly unparalleled in measurements focusing on the evolutionary rate of single components.

**METHODS:**

**Databases:** For a list of experimentally detected protein-protein interactions in *S. cerevisiae* we used the manually curated DIP database[9] (as of March 2003), that contains 3183 proteins with 9463 interactions. We assigned to each protein its known functional classification according to the MIPS database[15], which compiles genetic-, biochemical- and cell biological knowledge of yeast genes and proteins extracted from the literature. If a protein belongs to more than one functional class, its corresponding motif is assigned to both groups.

**Motif identification:** Similar to the method of detecting all *n*-node subgraphs of Milo *et al.*[7], our algorithm scans all rows of the adjacency matrix *M*. For each non-zero element *(i,j)* representing a link, it scans through all neighbors of *(i,j)*, $M_{ik}$, $M_{ki}$, $M_{jk}$, $M_{kj}$ = 1. This is performed recursively for all other elements *(i,k),(k,i),(k,j)* and *(j,k)* until a specific *n*-node subgraph is detected. Subsequently, the detected subgraphs are compared to the subgraphs found in previous steps and eliminated if they are already in the database. Note, that in contrast to Ref. 7, where motifs are defined as over-represented subgraphs, here we use the terms motifs and subgraphs interchangeably.

**Assigning orthologs:** The InParanoid database[11] provides orthologous sequence cluster information between organism pairs of *S. cerevisiae* and *H. sapiens, D. melanogaster, M. musculus, C. elegans* and *A. thaliana.* In our study, only the core orthologous sequence pair of each cluster was chosen, providing a bootstrap value of 100%. Each yeast protein that is engaged in orthologous core pairs in a specific eukaryote was labeled accordingly. Therefore, 2,174 proteins were labeled to have orthologs in *H. sapiens*, 2,093 in *A. thaliana*, 1,696 in *C. elegans*, 1,674 in *M. musculus* and 1,958 in *D. melanogaster*, respectively. This detailed ortholog information was used for calculating the results depicted in Fig. 1. For the data presented in Table 1 and 2, we identified 678 yeast proteins, which have an ortholog in each of the five higher organisms, representing the cross-section of the orthologous sets derived for the five organisms.

**Random ortholog distribution:** As a negative control set, we selected 678 proteins randomly on the yeast protein interaction network, assigned them as random orthologs and determined again the number of specific yeast motifs that are fully conserved (i.e., each of their components belong to the random ortholog set). The random conservation rate of a motif with $n$ proteins is well approximated by $p^n$, where $p$ is the probability that a protein has an ortholog across all five higher eukaryotes, given by $p = 678/3128 = 0.216$. Indeed, $p^n$ gives 4.6%, 1.01%, 0.22% and 0.047% for the two, three, four and five node motifs, respectively, in agreement with the numbers shown in Table 1 for the random conservation rate.

**Enrichment of Orthologous Proteins:** Starting from an orthologous protein $i$ we identify all proteins that are at distance $d$ from this protein, denoting their number with $N(d)$. For example, $N(1)$ would give the number of proteins directly interacting with protein $i$. Of the $N(d)$ proteins we also identify the number of proteins $n(d)$ that have an ortholog in a reference eukaryote. The ratio $r(d) = n(d)/N(d)$ gives the fraction of orthologs at distance $d$ from protein $i$. If the orthologs are randomly placed on the network, this ratio should be independent from $d$ and have the value $r = n/N$, where $n$ is the total number of yeast orthologs in the reference organism and $N$ is the total number of proteins in the network. The ratio $E(d) = r(d)/r$ gives the orthologous enrichment, which is equal to 1 for any $d$ if there is no clustering of orthologs in the network. $r(d) >> 1$ implies, that among proteins at distance $d$ from $i$ the orthologs are overrepresented, providing a signature of clustering. To decrease the noise level in Fig. 1d, $r(d)$ was averaged over all yeast orthologs chosen as $i$.

**Functional classes:** We determined the number of motifs ($\mu_h$) for the subnetworks defined by yeast proteins belonging to a specific functional class and found to be fully conserved in humans. In order to uncover overrepresented motifs in each functional class, we determined the average number of each motif ($\mu_r$) and the respective standard deviation ($\sigma_r$) using 100 random human ortholog sets. The parameter $Z = (\mu_h - \mu_r)/\sigma_r$ offers a quantitative measure of the degree to which a motif is overrepresented in a specific functional class: $Z >> 1$ implies that we find significantly more motifs in that class than a random distribution of ortholog placement could support. For functional classification we used the MIPS database[15], which classifies Yeast proteins in 17 distinct functional classes. This coarser classification offers better statistics for most classes.

**Clustering:** To characterize the degree of clustering in the network (Fig. 1c) we used the clustering coefficient, defined as $C_i = 2n_i/k_i(k_i-1)$, where $n_i$ is the number of direct links between the $k_i$ neighbors of protein $i$ [12]. The clustering coefficient is one if all neighbors of node $i$ are connected to each other while it is zero if none of the neighbors have links to each other.

# References

1. Hasty, J., McMillen, D. & Collins, J.J. Engineered gene circuits. *Nature* **420**, 224-230 (2002).

2. Kitano, H. Systems biology: a brief overview. *Science* **295**, 1662-1664 (2002).

3. Rao, C.V., Wolf, D.M. & Arkin, A.P. Control, exploitation and tolerance of intracellular noise. *Nature* **420**, 231-237 (2002).

4. Hartwell, L.H., Hopfield, J.J., Leibler, S. & Murray, A.W. From molecular to modular cell biology. *Nature* **402**, C47-52 (1999).

5. Oltvai, Z.N. & Barabási, A.-L. Life's complexity pyramid. *Science* **298**, 763-764 (2002).

6. Shen-Orr, S.S., Milo, R., Mangan, S. & Alon, U. Network motifs in the transcriptional regulation network of *Escherichia coli*. *Nat. Genet.* **31**, 64-68 (2002).

7. Milo, R., Shen-Orr, S.S., Itzkovitz, S., Kashtan, N. & Alon, U. Network motifs: simple building blocks of complex networks. *Science* **298**, 824-827 (2002).

8. Lee, T.I. *et al.* Transcriptional regulatory networks in *Saccharomyces cerevisiae*. *Science* **298**, 799-804 (2002).

9. Xenarios, I. *et al.* DIP, the Database of Interacting Proteins: a research tool for studying cellular networks of protein interactions. *Nucl. Acids Res.* **30**, 303-305 (2002).

10. Von Mering, C. *et al.* Comparative assessment of large-scale data sets of protein protein interactions. *Nature* **417**, 399-403 (2002).

11. Remm, M., Storm, C.E.V. & Sonnhammer, E.L. Automatic clustering of orthologs and in-paralogs from pairwise species comparisons. *J. Mol. Biol.* **314**, 1041-1052 (2001).

12. Watts, D.J. & Strogatz, S.H. Collective dynamics of 'small-world' networks. *Nature* **393**, 440-442 (1998).

13. Ravasz, E., Somera, A.L., Mongru, D. A., Oltvai, Z.N. & Barabási, A.-L. Hierarchical organization of modularity in metabolic networks. *Science* **297**, 1551-1555 (2002).

14. Rives, A.W. & Galitski, T. Modular organization of cellular networks. *Proc. Natl. Acad. Sci. U.S.A.* **100**, 1128-1133 (2003).

15. Mewes, H.W. et al. MIPS: a database for genomes and protein sequences. *Nucl. Acids Res.* **30**, 31-34 (2002).

16. Hurst, L.D. & Smith, N.G. Do essential genes evolve slowly? *Curr. Biol.* **9**, 747-750 (1999).

17. Hirsh, A.E. & Fraser, H.B. Protein dispensability and rate of evolution. *Nature* **411**, 1046-1049 (2001).

18. Hirsh, A.E. & Fraser, H.B. Genomic function (communication arising): Rate of evolution and gene dispensability. *Nature* **421**, 497-498 (2003).

19. Fraser, H.B., Hirsh, A.E., Steinmetz, L.M., Scharfe, C. & Feldman, M.W. Evolutionary rate in the protein interaction network. *Science* **296**, 750-752 (2002).

20. Pal, C., Papp, B. & Hurst, L.D. Genomic function (communication arising): Rate of evolution and gene dispensability. *Nature* **421**, 496-497 (2003).

21. Jordan, I.K., Rogozin, I.B., Wolf, Y.I. & Koonin, E.V. Essential genes are more evolutionarily conserved than are nonessential genes in bacteria. *Genome Res.* **12**, 962-968 (2002).

22. Snel, B., Bork, P. & Huynen, M.A. The identification of functional modules from the genomic association of genes. *Proc. Natl. Acad. Sci. U.S.A.* **99**, 5890-5895 (2002).

**Acknowledgement**
Research at the University of Notre Dame and Northwestern University was supported by grants from the National Institute of Health (NIGMS) and the Department of Energy Genomes to Life Program. Research at the Notre Dame was also supported by the National Science Foundation. Correspondence and request for materials should be sent to A.-L.B. (E-mail: alb@nd.edu) or Z.N.O. (E-mail: zno008@northwestern.edu).

| # | Motifs | Number of yeast motifs | Natural conservation rate | Random conservation rate | Conservation ratio |
|---|---|---|---|---|---|
| 1 | 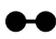 | 9,266 | 13.67% | 4.63% | 2.94 |
| 2 | 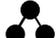 | 167,304 | 4.99% | 0.81% | 6.15 |
| 3 | 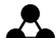 | 3,846 | 20.51% | 1.01% | 20.28 |
| 4 | 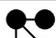 | 3,649,591 | 0.73% | 0.12% | 5.87 |
| 5 | 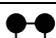 | 1,763,891 | 2.64% | 0.18% | 14.67 |
| 6 | 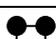 | 9,646 | 6.71% | 0.17% | 40.44 |
| 7 | 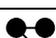 | 164,075 | 7.67% | 0.17% | 45.56 |
| 8 | 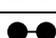 | 12,423 | 18.68% | 0.12% | 157.89 |
| 9 | 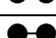 | 2,339 | 32.53% | 0.08% | 422.78 |
| 10 | 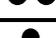 | 25,749 | 14.77% | 0.05% | 279.71 |
| 11 | 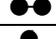 | 1,433 | 47.24% | 0.02% | 2,256.67 |

**Table 1: The evolutionary conservation of motif constituents.** The third column denotes the number of motifs of a given kind found in the yeast protein interaction network of 3,174 proteins, obtained by counting all subgraphs of two to five nodes (since there are 28 five-node motifs, we show only two (#10,11)). We identified 678 proteins that have an ortholog in each of the five studied higher eukaryotes, and identified all motifs for which each component belongs to this evolutionary conserved protein subset. The natural conservation rate shows what fraction of the original yeast motifs are evolutionary fully conserved, i.e., each of their protein components belong to the 678 orthologs of the list. For example, we find that 47% of the 1,433 fully connected pentagons (#11) found in yeast have each of their five proteins conserved in each of the five higher eukaryotes. If the topology of motifs does not interfere with the conservation rate of its constituting proteins, a random ortholog distribution should give the same conservation rate for specific motifs as seen in the natural sample (see Methods). The random conservation rate therefore denotes the fraction of motifs which are found to be fully conserved for the random ortholog distribution. The last column denotes the ratio between the natural and the random conservation ratios, indicating that all motifs are highly conserved, some (for example #11) having a natural conservation rate 2,256 times higher than expected in the absence of correlations between protein conservation rate and the topology of a given motif.

| Functional class | Overrepresented motifs |
|---|---|
| Transport facilitation | 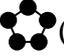(10) |
| Subcellular localization | 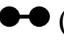(21) 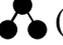(21) 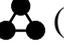(26) 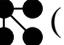(15) 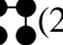(27) 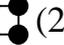(23) 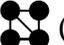(29) 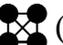(20) 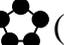(63) 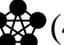(45) |
| Regulation | 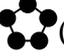(10) |
| Protein fate | 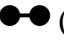(14) 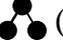(16) 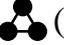(13) 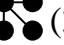(33) 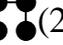(27) 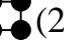(20) 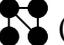(26) 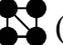(24) 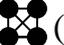(16) 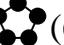(60) 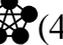(41) |
| Cell cycle | 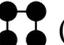(11) 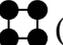(14) 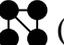(13) 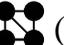(11) 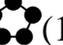(14) |
| Cellular transport | 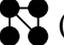(11) 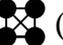(12) |
| Transcription | 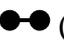(12) 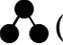(16) 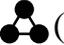(17) 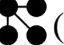(13) 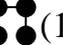(16) 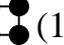(19) 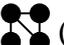(17) 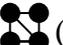(15) 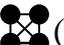(14) 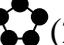(21) 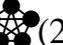(23) |
| Protein synthesis | 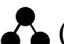(12) 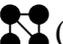(11) 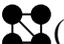(17) 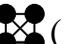(11) 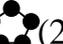(24) |

**Table 2: Overrepresentation of human orthologous motifs in various functional classes of yeast proteins.** We determined the number of motifs for the subnetworks defined by proteins belonging to a specific functional class, as well as the number of these motifs ($\mu_h$) that are fully conserved in humans. Finally, for 100 randomized human orthologous sets we determined the average number of motifs ($\mu_r$) in the random ortholog samples and the standart deviation ($\sigma_r$) for each motif. The table lists all motifs that are at least ten times overrepresented compared to a random configuration ($Z > 10$, see Methods), the specific $Z$ values being shown next to the motifs. We did not find overrepresented motifs for the classes of transposable elements, energy, cellular fate, cellular communication, cellular rescue, cellular organization, metabolism, protein activity, protein binding and proteins which are not classified yet or classified unclearly. Note, that if all proteins of a given motif simultaneously belong to more than one functional class, the motif will also appear in multiple functional classes.

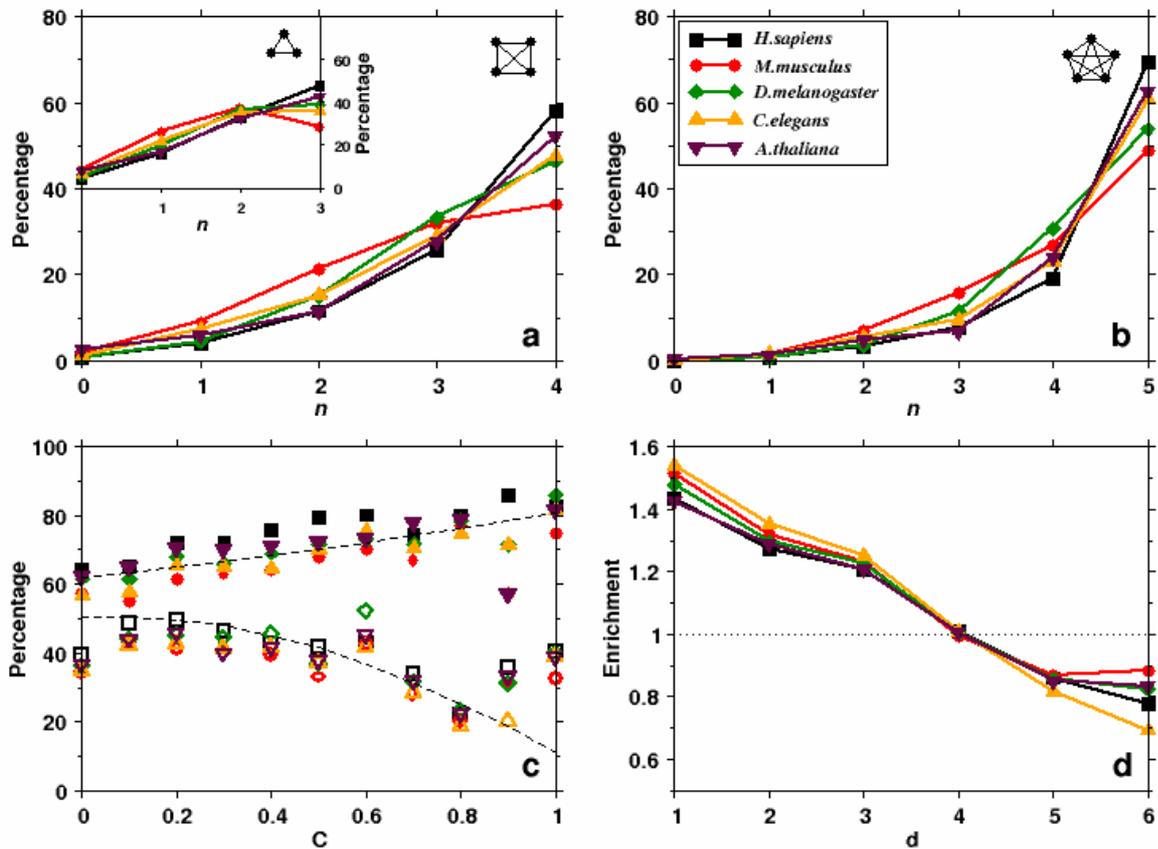

**Figure 1: Relationship between the topology of a protein interaction network and the evolutionary conservation of individual proteins.** Panel (a,b) shows the detailed conservation rates of fully connected 3- (inset a), 4- (a) and 5- (b) node motifs. For example, (b) indicates that in humans less than 1% of the 1,433 pentagon motifs found originally in yeast have fully disappeared (i.e., none of their components have an ortholog), and only 1.5% of motifs have a single ortholog component ($n = 1$), while for more than 69% of the motifs each of the five proteins have been conserved ($n = 5$). The five curves correspond to the five studied eukaryotes, the legend in (b) identifying the corresponding symbols and colors used throughout (a-d). (c) The conserved fraction of the immediate neighbors of an orthologous protein $i$ (filled symbols) correlates positively with the node's clustering coefficient $C$. Open symbols show the fraction of orthologs in the vicinity of a non-ortholog protein displaying a negative trend with $C$. (d) The enrichment, defined as the ratio between the percentage of orthologous proteins at distance $d$ from an ortholog in the natural- and the random orthologous sets, indicates decreasing overrepresentation of orthologs with increasing distance.

16